# Evidence for *d*-wave superconductivity in single layer FeSe/SrTiO$_3$ probed by quasiparticle scattering off step edges

Zhuozhi Ge,[†,‡] Chenhui Yan,[†] Huimin Zhang,[†] Daniel Agterberg,[‡] Michael Weinert,[‡] and Lian Li[*,†]

[†]Department of Physics and Astronomy, West Virginia University, Morgantown, WV 26506, USA

[‡]Department of Physics, University of Wisconsin, Milwaukee, WI 53211, USA

*Email: lian.li@mail.wvu.edu

Phone: (+1) 304-293-4270

ABSTRACT: The de Gennes extrapolation length is a direction dependent measure of the spatial evolution of the pairing gap near the boundary of a superconductor, and thus provides a viable means to probe its symmetry. It is expected to be infinite and isotropic for plain *s*-wave pairing, and finite and anisotropic for *d*-wave. Here, we synthesize single layer FeSe films on $SrTiO_3(001)$ (STO) substrates by molecular beam epitaxy, and measure the de Gennes extrapolation length by scanning tunneling microscopy/spectroscopy. We find a 40% reduction of the superconducting gap near specular $[110]_{Fe}$ edges, yielding an extrapolation length of 8.0 nm. However, near specular $[010]_{Fe}$ edges the extrapolation length is nearly infinite. These findings are consistent with a phase changing pairing with 2-fold symmetry, indicating *d*-wave superconductivity. This is further supported by the presence of in-gap states near the specular $[110]_{Fe}$ edges, but not the $[010]_{Fe}$ edges. This work provides direct experimental evidence for *d*-wave superconductivity in single layer FeSe/STO, and demonstrates quasiparticle scattering at boundaries to be a viable phase sensitive probe of pairing symmetry in Fe-based superconductors.

A central milestone in the search for high temperature Fe-based superconductors (FeSCs)[1,2] is the determination of pairing gap symmetry.[3-6] Currently, the commonly presumed gap structure for FeSCs is sign reversing *s+-* pairing, which results from interband scattering between hole pockets around Γ point and electron pockets around M point in the Brillouin zone (BZ).[3,7] This mechanism, however, presents a conundrum in the recently discovered single layer FeSe/SrTiO$_3$ (STO) with the highest superconducting temperature ($T_C$) to date amongst all FeSCs.[8-10] Due to charge doping from the substrate, the Fermi surface of FeSe consists of only electron pockets centered at M, with the states at Γ completely pushed below the Fermi level.[11-14] Clearly, this poses a challenge for pairing theories that involve both pockets.[3,9] While plain *s*-wave pairing was suggested based on earlier angle-resolved photoemission spectroscopy (ARPES) observations and scanning tunneling microscopy/spectroscopy (STM/S) measurements of quasiparticle interference (QPI),[11,15] recent ARPES measurements indicate gap anisotropy.[16] This anisotropy is naturally explained by a nodeless *d*-wave state, for which the observed gap minima are a manifestation of nodes that have not formed.[17] However, gap anisotropy is not a robust probe of pairing symmetry and direct confirmation of any pairing symmetry by phase sensitive measurements is sorely needed.[3]

One such approach is corner junction measurements,[18,19] which have revealed *d*-wave pairing in YBa$_2$Cu$_3$O$_{7-\delta}$. However, this is challenging for the single layer FeSe system, because the interfacial Josephson current will be suppressed for all states except a plain *s*-wave state due to gaps of opposite signs on different bands. Alternatively, QPI measurements using STM can provide phase sensitive information, however, if the atomic impurities are located at the top surface atomic plane, such scattering may not provide sufficient information on the pairing symmetry.[9,20] Furthermore, recent model calculations of QPI patterns near a magnetic impurity in single layer

FeSe indicate that when the spin-orbit coupling is smaller than the superconducting gap, they appear as *s*-wave, even though the pairing potential is *d*-wave.[21]

Within weak-coupling theory,[22] quasiparticle scattering at the boundary of a superconductor is determined by the symmetry of the order parameter (OP) and hence provides another avenue to probe the pairing symmetry.[23] For example, a superconducting OP with phase changes can lead to antiphase interference and suppress superconductivity at the boundary for certain orientations.[23] This effect can be described by the de Gennes extrapolation length $b$,[24] which reflects the spatial evolution of the OP near the boundary (Figure 1a).[25] For example, for elastic scattering off a specular edge in two-dimensions, the ***q*** vector is always perpendicular to the edge (see Supporting Information Figure S1). Thus, $b$ depends critically on the orientation of the boundary for *d*-wave (Figures S1b and S1c), but not for *s*-wave superconductors (Figures S1d and S1e), providing a definitive signature for anisotropic pairing. In the case of single layer FeSe/STO,[9] for *d*-wave pairing (Figure 1c),[17,26,27] the OP will change sign after a 90° rotation, leading to destructive interference that reduces the pairing gap near the boundary. On the other hand, for *s*-wave (Figure 1d),[28-31] the phase of the OP is preserved after a 90° rotation, resulting in a constant gap near the boundary. An additional consequence for *d*-wave pairing is the appearance of Andreev bound states within the gap.[32-35] As summarized in Table 1, the measurements of $b$ and in-gap states near different edges can provide a clear indication of the paring symmetry in single layer FeSe/STO.

Here, we present a systematic investigation of edge scattering in superconducting single layer FeSe/STO. Two types of films were grown by MBE in an ultrahigh vacuum system with a base pressure below $1.0\times10^{-10}$ Torr (see Supporting Information for experimental methods). The as-grown films (see Supporting Information Figure S2a) are not superconducting due to excess Se in Se-rich growth conditions.[8] They were then extensively annealed to induce superconductivity, as verified by *in situ* STM/S and ARPES measurements (see Supporting Information Figures S3 and S5). Furthermore, different annealing conditions were used to produce different types of step edges, as discussed in more details below.

In the first case, single layer FeSe films (Figure 2a) exhibit two types of edges: edges of pits, and those at the low-contrast trench grain boundaries that consist of a missing Fe row and a shift of half unit cell in the perpendicular direction.[36,37] Figure 2b is an atomic resolution image of the specular edges (also see Supporting Information Figure S2b). Note that while the edges exhibit atomic scale disorders on the order of ~0.4 nm, they are one order of magnitude smaller than the superconducting coherence length of ~3.2 nm,[15] and hence can still be considered as specular edges. The square arrays of bright features correspond to surface Se atoms. The Fe square lattice lies in the layer below and is diagonally aligned to the surface Se along the labeled *x/y* directions (Figure 1b). Thus, both the edges along the pits and grain boundaries are specular $[110]_{Fe}$ edges. Away from the edges, a superconducting gap of $19.0 \pm 0.8$ meV is found by dI/dV spectroscopy (Figure S3a). A superconducting temperature of 49 K is determined from the BCS-like fitting of temperature dependent gaps (Figure S3c), close to our ARPES measurements (Figure S5) and earlier work.[8,36]

In the second case, the FeSe films exhibit randomly oriented high-contrast grain boundaries, as shown in Figure 2c, where there is a 1/2 unit cell offset along the $[010]_{Fe}$ direction but no missing atom rows (see Supporting Information Figure S6). This difference in topography is likely due to different post-annealing conditions, and/or the FeSe/STO interfaces.[15] The superconducting gap of this type of sample is also slightly smaller than that of FeSe #1 (Figure S3a), but within the variation from sample to sample that is typically observed. Interestingly, specular $[010]_{Fe}$ edges are formed along the steps, shown in Figure 2d, with an atomic resolution image showing the Se square lattice rotated 45° from the *x/y* directions. Rough [010]$_{Fe}$ edges are also present in this type of samples (Figure S2c).

To determine the extrapolation length, spatially resolved dI/dV spectra were taken on both the specular [110]$_{Fe}$ edge on FeSe #1 (Figures 3a and 3c, from pit edge) and the specular [010]$_{Fe}$ edge on FeSe #2 (Figures 3b and 3d, from step edge) (The large scale images where Figures 3a and 3b are taken are shown in Supporting Information Figure S7). For the specular [110]$_{Fe}$ edge, dI/dV spectra taken along the black arrow are shown in Figure 3c. A superconducting gap is observed for all spectra (see Supporting Information Figure S4 for procedure to determine the gap). However, the gap magnitude varies: it is nearly constant at 18.0 meV for the first 3.5 nm, and then monotonically decreases to 12.8 meV towards the edge. We note that this reduction in the superconducting gap is constant along the edge, as shown in Supporting Information Figure S8, indicating that suppression in superconductivity is not a local effect. The spatial dependence of the superconducting gap is summarized in Figure 3e, which yields an extrapolation length of 8.0 nm. For other [110]$_{Fe}$ edges in both types of FeSe films, similar analysis reveals a *b* of 7.8 nm near trench grain boundary (see Supporting Information Figure S9), and 9.6 nm near step edges (see Supporting Information Figure S10). Note that the dI/dV spectra shown here exhibit only one-gap structure, likely due to thermal broadening. Measurements were also carried out in a separate STM (Unisoku 1300) at a lower temperature of 4 K, where two-gap features are resolved (see Supporting Information Figure S11a). A similar reduction of the superconducting gap is also observed for the inner gap, while the outer gap remains constant. An extrapolation length *b* of 8.4 nm is obtained near a trench grain boundary (Figure S11c).

The behavior is quite different for the specular [010]$_{Fe}$ edges (Figure 3b). As shown by the spatially resolved dI/dV spectra taken along the blue arrow from point 1 to point 18 in Figure 3d, the gap size is nearly constant at 13.9 ± 0.8 meV at all positions, indicating a nearly infinite extrapolation length (Figure 3e). Similar behavior is reproducible on other specular [010]$_{Fe}$ edges (see Supporting Information Figure S12).

This orientation dependent extrapolation length suggests a 2-fold anisotropy in the superconducting order parameter, which is expected for *d*-wave pairing, but not for *s*-wave. For $d_{x^2-y^2}$ wave (Figure 1c), the OP changes sign under 90° rotation, and exhibits a symmetry such that $\Delta(\boldsymbol{k}_{in}) = -\Delta(\boldsymbol{k}_{out})$ for [110]$_{Fe}$ edge scattering, where $\boldsymbol{k}_{in}$ and $\boldsymbol{k}_{out}$ are the incoming and reflected wave vectors. The antiphase interference between $\Delta(\boldsymbol{k}_{in})$ and $\Delta(\boldsymbol{k}_{out})$ will reduce the order parameter and suppress superconductivity near the [110]$_{Fe}$ edge. For the [010]$_{Fe}$ edge scattering, the OP has a symmetry that warrants $\Delta(\boldsymbol{k}_{in}) = \Delta(\boldsymbol{k}_{out})$, thus will not change the pairing gap. Hence no pair breaking effect is expected at the [010]$_{Fe}$ edge for *d*-wave pairing.

On the other hand, for plain *s*-wave (Figure 1d) and other types of *s*-wave pairing (Figures 1e and 1f), the phase of the order parameter is preserved under 90° rotation, thus $\Delta(\boldsymbol{k}_{in}) = \Delta(\boldsymbol{k}_{out})$ for both the [110]$_{Fe}$ and [010]$_{Fe}$ edge scatterings. The *s*-wave superconducting gap will be constant and the extrapolation length will be infinite regardless of the edge orientation. This is consistent with earlier observations of nearly constant superconducting gaps near the edge of Pb islands with conventional *s*-wave pairing.[38] For bonding-antibonding *s* wave pairing (Figure 1f), the OP has two phases around the M point, hence antiphase interference can, in principle, occur via interband scattering between the electron pockets. However, here the pair breaking effect is independent of edge orientations, and would occur on both specular [110]$_{Fe}$ and [010]$_{Fe}$ edges, inconsistent with our observations of large anisotropic extrapolation length (Figure 3).

We note that the weak-coupling theory previously developed for the extrapolation length is for single band systems,[22] while single layer FeSe is multiband.[17] This feature does not usually qualitatively alter the theory. However, the fact that two bands have energy differences less than the gap energy in some momentum-space directions is a new feature and gives rise to an interband pairing that leads to the nodeless gap.[17] In the calculations of the extrapolation length, such an interband is not considered and may play a role a role in FeSe. However, once the band splitting is greater than the gap energy, only intraband pairing is important and the usual theory can be used. This applies to most of the Fermi surface of single layer FeSe, except a narrow range near where the nodes should have been. Hence the nodeless *d*-wave order parameter will have sign changes as in earlier theories and the pairing symmetry analysis from the extrapolation length can be applied to our experiment.

Furthermore, the fact that $b$ is much larger than coherence length (~3.2 nm[15]) can also be explained by the nodeless *d*-wave pairing model.[17] Here the interband interaction in the region near the node opening is responsible for the nodelessness and should be considered as a correction to the intraband interaction. It weakens the overall intraband antiphase interference,[23] resulting in weaker pair breaking and hence longer extrapolation length.

As discussed above, a nodeless *d*-wave pairing can also give rise to in-gap states at finite energies for specular $[110]_{Fe}$ edges, in contrast to the usually expected zero energy states for a nodal *d*-wave superconductor.[32,39] Single layer FeSe has a multiorbital Fermi surface. On the $[110]_{Fe}$ edges, orbital mismatch during the scattering leads to the formation of in-gap states which are not at the Fermi level.[16,32] This is confirmed by the presence of a finite energy bound state in the numerical energy spectra for this orientation.[39] (Note that following the theory of Ref. 32, no in-gap Andreev bound states are found for $[100]_{Fe}$ edges due to the nearly orthogonal orbitals on the two different bands in this momentum space direction). To better visualize the appearance of in-gap states, we normalize the spatially resolved dI/dV spectra shown in Figures. 3c and 3d by subtracting spectrum 1 (away from the edge), as shown in Figures 4a and 4b. For the specular $[110]_{Fe}$ edge, in-gap states (peaked between -6 and 8 meV) appear at ~ 4 nm from the edge and enhances significantly towards the edge. Note that the peak positions are constant, as marked by the two dashed lines in Figure 4a, suggesting the in-gap states are inherent (see Supporting Information Figure S13 for more details). In contrast, for the specular $[010]_{Fe}$ edge, only weak fluctuations are observed in the normalized dI/dV spectra with no in-gap states present. Comparison to the expected signatures of several proposed pairing symmetries in Table 1, our findings indicate *d*-wave pairing for single layer FeSe/STO.

The pair breaking effect should also be sensitive to the roughness of the edge.[23] In the presence of nanoscale disorder, the scattering angle cannot be precisely defined as it will be a mixture of several possible directions including $[110]_{Fe}$. Spatially resolved dI/dV spectra taken near a rough $[010]_{Fe}$ edge on FeSe #2 is shown in Supporting Information Figure S14, yielding an extrapolation length of 16.3 nm. Similarly, in-gap states also appear. These results are in excellent agreement with that would be expected for scattering off a rough edge for a *d*-wave superconductor.

In conclusion, we have epitaxially grown superconducting single layer FeSe films on STO(001) substrates with various types of edges. For the specular $[110]_{Fe}$ edge, spatially resolved dI/dV spectroscopy reveals a suppression of superconductivity near the edge with an extrapolation length of 8.0 nm. In contrast, at the specular $[010]_{Fe}$ edge, no suppression of superconductivity is observed with a near infinite extrapolation length. This edge orientation dependence on Cooper pairing destabilization is consistent with nodeless *d*-wave pairing symmetry, which is further supported by the presence of in-gap states near the specular [110]$_{Fe}$ edges, but not the [010]$_{Fe}$ edges. Our findings further demonstrate that quasiparticle scattering at the boundaries of nanostructures is a viable phase sensitive probe of pairing symmetry of Fe-based superconductors.

**Table 1. Expected extrapolation length *b* and in-gap states (IGS) at different edges of single layer FeSe for different pairing symmetries.**

|  | *d* wave | Plain *s* wave | Incipient *s* wave | Bond-antibonding *s* wave |
|---|---|---|---|---|
| *b* at $[110]_{Fe}$ | Finite | Infinite | Infinite | Finite |
| IGS at $[110]_{Fe}$ | Yes | No | No | Yes |
| *b* at $[010]_{Fe}$ | Infinite | Infinite | Infinite | Finite |
| IGS at $[010]_{Fe}$ | No | No | No | Yes |

FIGURES

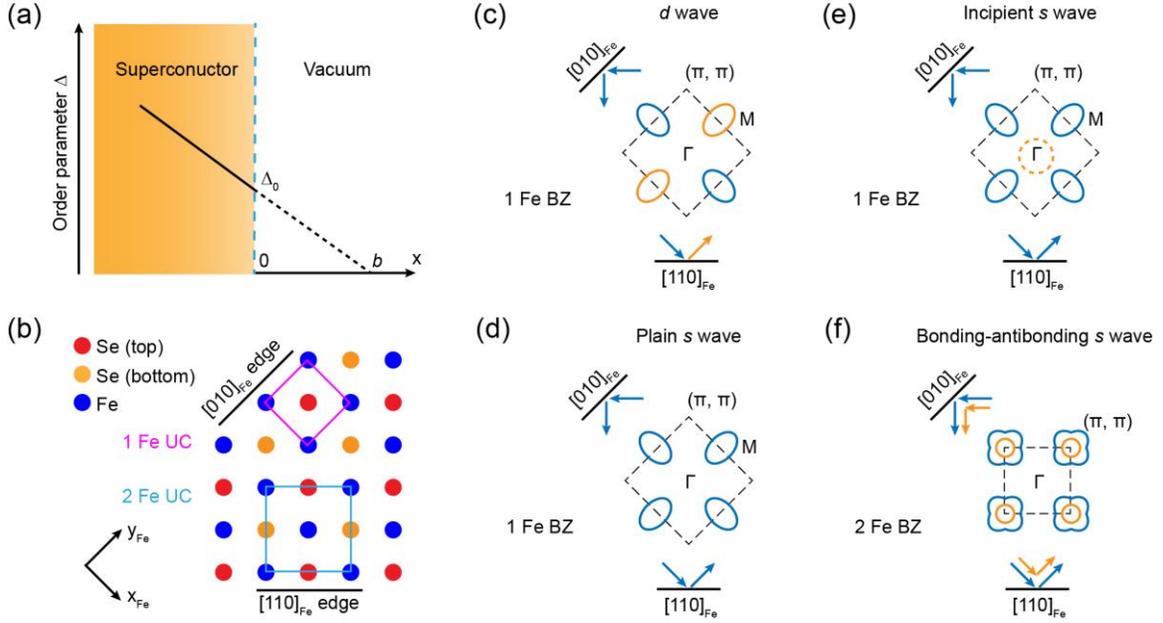

**Figure 1.** (a) Schematic diagram of the de Gennes extrapolation length. The inclined solid black line shows the reduction of the superconducting OP near the boundary. $\Delta_0$ is the OP at the boundary. The inclined dashed black line demonstrates linear extrapolation of the OP into the vacuum, and intersection of the *x*-axis is the extrapolation length *b*. (b) Crystal structure of single layer FeSe. The $[010]_{Fe}$ and $[110]_{Fe}$ edges are defined in the Fe plane. Both one Fe unit cell (1 Fe UC) and two Fe unit cell (2 Fe UC) are shown in different colors. (c-f) *d* wave, plain *s* wave, and incipient *s* wave pairing in the 1 Fe BZ, and bonding-antibonding *s* wave pairing in the 2 Fe BZ. The ellipses represent the gap structure and the arrows the quasiparticle scattering off the $[010]_{Fe}$ and $[110]_{Fe}$ edges. Blue indicates positive and yellow negative OP.

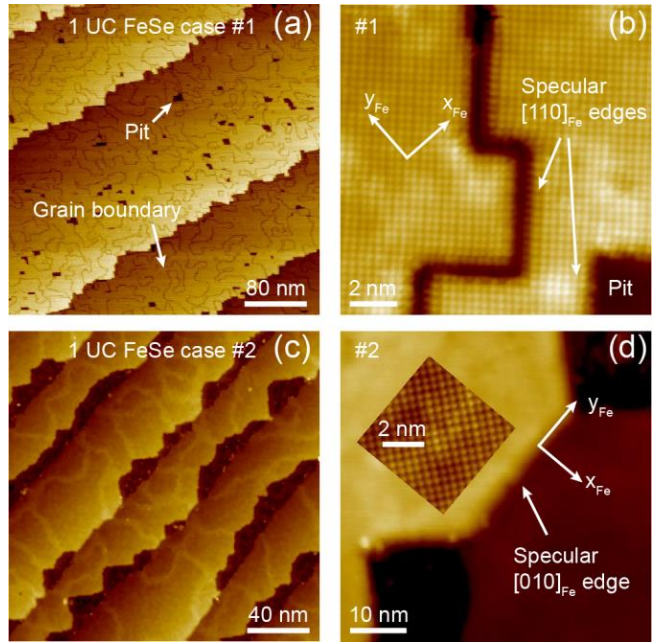

**Figure 2.** (a) STM image of annealed single layer (1 UC) FeSe/STO film #1 ($V_s$= 1.0 V, $I_t$= 0.1 nA). (b) Atomic resolution image of specular $[110]_{Fe}$ edges on FeSe film #1 ($V_s$= 0.5 V, $I_t$= 0.1 nA). Two types of $[110]_{Fe}$ edges are marked: at the trench grain boundary and at the FeSe-to-STO edge. (c) STM image of annealed 1UC FeSe film #2 ($V_s$= 1.2 V, $I_t$= 0.1 nA). (d) STM image of a specular $[010]_{Fe}$ step edge on FeSe #2 ($V_s$= 1.0 V, $I_t$= 0.1 nA). Inset is an atomic resolution image of the Se lattice ($V_s$= 0.5 V, $I_t$= 0.1 nA).

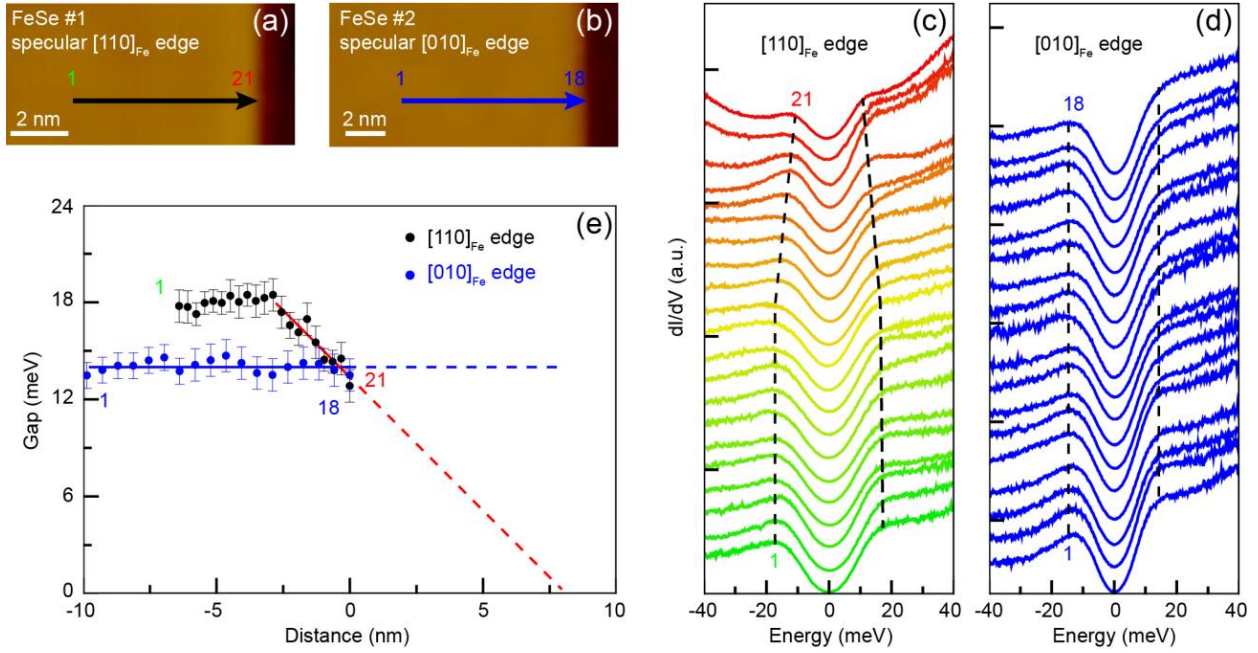

**Figure 3.** (a) STM image of a specular $[110]_{Fe}$ edge ($V_s$= 1.0 V, $I_t$= 0.1 nA). (b) STM image of a specular $[010]_{Fe}$ edge ($V_s$= 1.0 V, $I_t$= 0.1 nA). (c) Spatially resolved dI/dV spectra taken perpendicular to the $[110]_{Fe}$ edge, along the black arrow in (a). The dashed black lines are guide to the evolution of the gap. (d) Spatially resolved dI/dV spectra taken perpendicular to the $[010]_{Fe}$ edge, along the blue arrow in (b). (e) Profile of the measured superconducting gaps perpendicular to the $[110]_{Fe}$ (black dots) and $[010]_{Fe}$ (blue dots) edges. The solid red line is a linear fit of the gaps and the dashed red line is the linear extrapolation. The fitting function is (13.3-1.67$x$) for the $[110]_{Fe}$ edge, where $x$ is the distance from the edge and the origin of $x$ is right on the edge. The solid blue line marks the average gap perpendicular to the $[110]_{Fe}$ edge and the dashed blue line is the linear extrapolation.

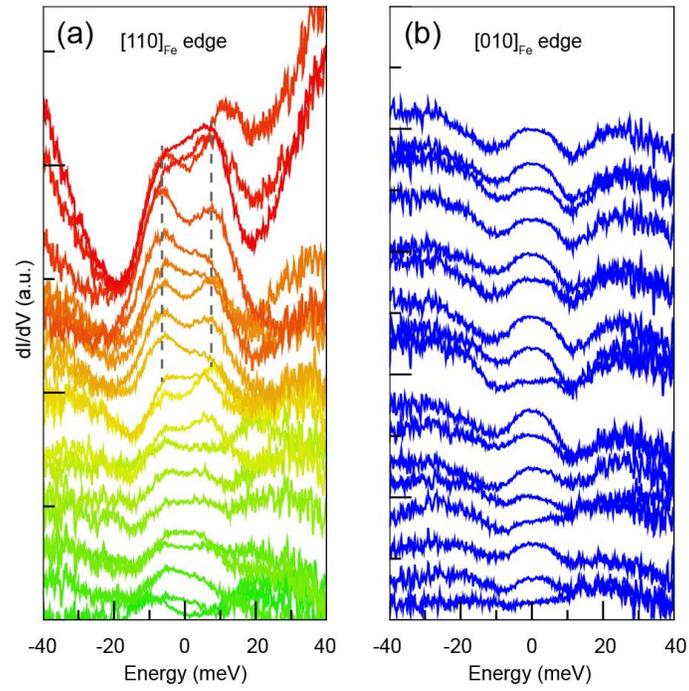

**Figure 4.** (a) The same spatially resolved dI/dV spectra as in Figure 3c, normalized by subtracting the spectrum 1. The two dashed dark lines mark the peaked in-gap states. (b) The same spatially resolved dI/dV spectra as in Figure 3d, normalized by subtracting the spectrum 1.

AUTHOR INFORMATION

**Corresponding Author**

*E-mail: lian.li@mail.wvu.edu.

**ORCID**

Zhuozhi Ge: 0000-0001-5937-4443

Lian Li: 0000-0001-5011-7774

**Author Contributions**

Z.G. and L.L. conceived and designed the experiments. Z.G. grew the FeSe films. Z.G. and C.Y. carried out the STM/S measurements. L.L. and Z.G. wrote the paper. All authors discussed the results and commented on the manuscript.

**Notes**

The authors declare no competing interests.

ACKNOWLEDGMENT

We gratefully acknowledge funding by the U.S. National Science Foundation (DMR-1335215), and Department of Energy (DE-SC0017632).

TABLE OF CONTENTS

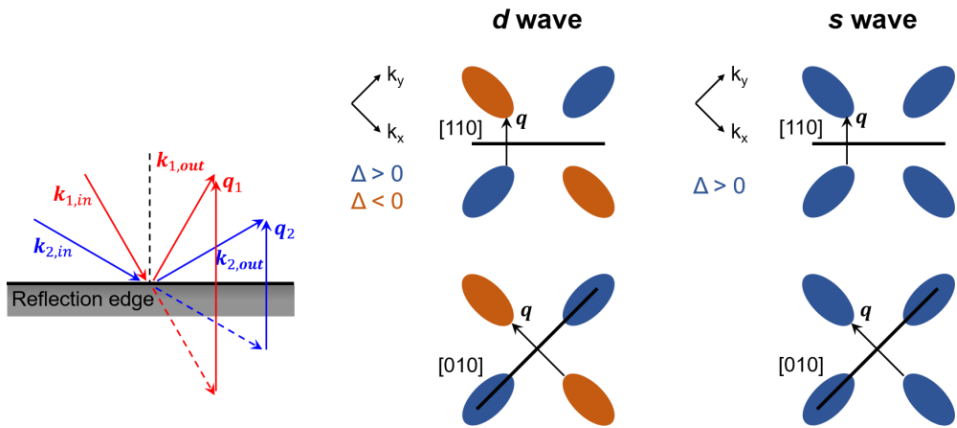